\documentclass{epl}

\title{Possibility of reflectionless tunneling crossed transport at
normal metal~/ superconductor double interfaces}
\shorttitle{Reflectionless tunneling ...}
\author{S. Duhot and R. M\'elin}
\institute{
Centre de Recherches sur les Tr\`es Basses
Temp\'eratures, CRTBT\cite{crtbt},\\ CNRS, BP 166,
38042 Grenoble Cedex 9, France
}


\begin{document}

\maketitle

\begin{abstract}
We investigate
one dimensional models (the Blonder,
Tinkham, Klapwijk model and a tight-binding model)
of non local transport at normal metal / superconductor (NS) double interfaces.
We find a negative elastic cotunneling
crossed conductance, strongly enhanced
by additional scatterers away from the interfaces,
suggesting the possibility of reflectionless tunneling
non local transport at double NS
interfaces with contacts having a sufficiently small extension.
\end{abstract}

\section{Introduction}

Single electron tunneling in a superconductor is prohibited
if the applied bias
voltage is smaller than the superconducting gap. However, an electron
in the spin-up band can be reflected as a hole in the spin-down band,
a phenomenon called Andreev reflection \cite{Andreev}.
A charge $2e$ is transmitted in the superconductor at each
Andreev reflection, so that the conductance of a
highly transparent 
normal metal / superconductor (NS) contact is doubled compared
to the one of the corresponding NN contact. The equilibrium
properties of the superconductor (such as the value of the self-consistent
superconducting gap) are modified by a normal electrode
connected to the superconductor, a phenomenon called the inverse
proximity effect. It
is expected that most of the inverse proximity effect takes place on
a length $a$
if the area of the contact $a^2$ is much smaller than
the superconducting coherence length $\xi$ \cite{BTK}.
The influence of the inverse proximity effect on 
transport properties can then be neglected, and a single channel,
ballistic, one-dimensional model with a step-function variation
of the superconducting gap captures the essential physics
of localized interfaces, as shown
by Blonder, Tinkham, and Klapwijk (BTK) \cite{BTK}. Moreover,
BTK introduce a repulsive potential at the NS interface, characterized
by the dimensionless parameter $Z$, being the 
strength of the repulsive potential normalized to the Fermi energy.
Transparent interfaces correspond to $Z=0$ and tunnel interfaces
correspond to $Z \gg 1$.

Disorder in the normal metal modifies strongly
subgap transport at a single normal metal~/ insulator
/ superconductor (NIS) interface \cite{Hekking-Nazarov,Beenakker-revue}.
The conductance can be enhanced by orders of magnitude
by constructive interferences in which
an electron can ``try'' the tunneling
process a huge number of times \cite{Hekking-Nazarov}.
This effect due to scattering by disorder is already present
in simple double barrier one-dimensional models. 
Melsen and Beenakker \cite{Melsen} consider a NINIS double
junction in one dimension,
with, in the BTK language, barrier parameters $Z_1$ (for the NIN interface)
and $Z_2$ (for the NIS interface). The conductance, averaged over
the Fermi oscillations, shows a maximum 
for a value of $Z_1$ comparable to $Z_2$
as $Z_1$ increases
while $Z_2 \gg 1$ is fixed \cite{Melsen}. 
This enhancement of the conductance shows that
the double barrier model captures
multiple scattering as in a disordered system.

\begin{figure}
\begin{center}
\includegraphics [width=.7 \linewidth]{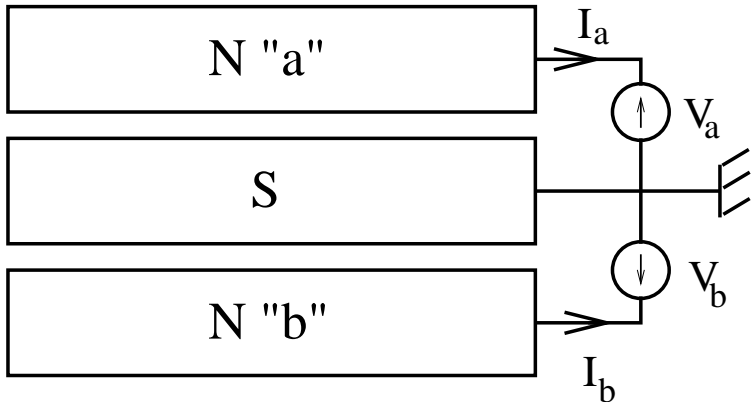}
\end{center}
\caption{Schematic representation of the electrical circuit corresponding to 
the NISIN junction studied experimentally
in Ref. \cite{Russo}. The current $I_a$ through electrode
``a'' is determined in response to a voltage $V_b$ on electrode ``b'',
with $V_a=0$.
\label{fig:schema-NISIN}
}
\end{figure}

We address here similar effects 
for non local transport in NISIN junctions
\cite{Lambert,Jedema,Byers,Deutscher,Samuelsson,Prada,Koltai,Falci,Melin-Peysson,Chte,japs,Melin-Feinberg,Tadei,Pistol,Melin}
in which the 
normal electrode ``a'' is at potential $V_a$, the 
electrode ``b'' is at potential $V_b$, and the superconductor
is at potential $V_S$ (see Fig.~\ref{fig:schema-NISIN}).
The non local conductance 
${\cal G}_{a,b}(V_b)$ contains the
information on how the current $I_a(V_b)$
in electrode ``a'' depends on the voltage $V_b$ on electrode
``b'': ${\cal G}_{a,b}(V_b)=\partial I_a(V_b)/\partial V_b$
\cite{Deutscher,Falci}.
The superconductor is taken as the reference voltage ($V_S=0$), and
we focus on the case $V_a=0$. Such devices have been realized
in two recent experiments, performed in Karlsruhe by Beckmann 
{\it et al.} with ferromagnets
\cite{Beckmann}, and in Delft by Russo {\it et al.}
with a NISIN trilayer \cite{Russo}. A sizeable crossed signal
is measured in the latter \cite{Russo}, which is surprising in view of 
lowest order perturbation theory in the tunnel
amplitudes predicting an exact cancellation between the
electron-electron and
electron-hole channel
crossed conductances \cite{Falci}. We take as a working hypothesis that
non local transport with normal metals
is described by higher order contributions in perturbation
theory in the tunnel amplitudes.
These were already evaluated in Ref. \cite{Melin-Feinberg}
within microscopic Green's functions
for localized interfaces. This approach was continued in Ref.~\cite{Melin} 
to account for extended interfaces with a large normal metal phase
coherence length, giving rise to weak localization.
Our task here is to investigate related issues in simple one dimensional
models in the spirit of Ref. \cite{Melsen}.

\begin{figure}
\begin{center}
\includegraphics [width=.7 \linewidth]{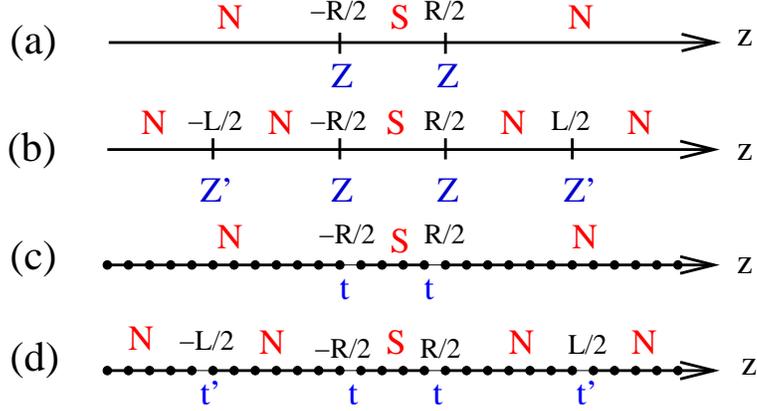}
\end{center}
\caption{(Color online) Schematic representations of the
one dimensional models: the BTK model for NISIN (a) and
NINISININ junctions (b), and the tight-binding model for
NISIN (c) and NINISININ (d) junctions.
\label{fig:schem-btk}
}
\end{figure}
\section{Blonder, Tinkham, Klapwijk (BTK) approach to a NISIN junction}
Let us first consider a one dimensional model of NISIN double
interface within the BTK approach \cite{japs} (see Fig.~\ref{fig:schem-btk}a).
The gap of the superconductor is supposed to have a step-function
variation: $\Delta(z)=\Delta \theta(z-R/2)
\theta(R/2-z)$, and we suppose $\delta$-function scattering potentials
at the interfaces:
$V(z)=H \delta(z+R/2)+H \delta(z-R/2)$
\cite{BTK}. The two-component wave-functions are given by
\begin{eqnarray}
 \psi_1(z) &=&
\left(\begin{array}{c}1\\0\end{array}\right) e^{i k_F z}
+a \left(\begin{array}{c}0\\1\end{array}\right) e^{i k_F z}
+b \left(\begin{array}{c}1\\0\end{array}\right) e^{-i k_F z}\\
\psi_2(z) &=&
c \left(\begin{array}{c}u_0\\v_0\end{array}\right) e^{i k_F (z+R/2)}
e^{-(z+R/2)/\xi}
+ d \left(\begin{array}{c}v_0\\u_0\end{array}\right) e^{-i k_F (z+R/2)}
e^{-(z+R/2)/\xi}\\
\nonumber
&+&c' \left(\begin{array}{c}u_0\\v_0\end{array}\right) e^{-i k_F (z-R/2)}
e^{(z-R/2)/\xi}
+d' \left(\begin{array}{c}v_0\\u_0\end{array}\right) e^{i k_F (z-R/2)}
e^{(z-R/2)/\xi}\\
 \psi_3(z) &=&
a' \left(\begin{array}{c}0\\1\end{array}\right) e^{-i k_F (z-R/2)}
+b' \left(\begin{array}{c}1\\0\end{array}\right) e^{i k_F (z-R/2)}
,
\end{eqnarray}
where $\psi_1(z)$, $\psi_2(z)$ and $\psi_3(z)$ correspond
respectively to $z<-R/2$, $-R/2<z<R/2$ and $R/2<z$,
and $u_0^2=1-v_0^2=\left(1+i\sqrt{\Delta^2-\omega^2}/\omega
\right)/2$ are the BCS coherence factors.
We introduce the parameter $Z=2mH/\hbar^2 k_F$.
The unknown coefficients $a$, $b$, $a'$, $b'$,
$c$, $d$, $c'$, $d'$ are determined from
matching the wave-functions and their derivatives \cite{BTK}.
Assuming $R\gg \xi$,
we expand $a'$ and $b'$ to first order in $\exp{(-R/\xi)}$, to find
the transmission coefficients
\begin{eqnarray}
&&\int_0^{2\pi} \frac{d (k_F R)}{2 \pi} |a'(k_F R)|^2 =
 \left( \frac{1}{2 Z^4}-\frac{1}{2 Z^6}+\frac{1}{2 Z^8}
+ ...\right) \exp{(-2R/\xi)} 
+ {\cal O}\left(\exp{(-4R/\xi)}\right)\\
&&\int_0^{2\pi} \frac{d (k_F R)}{2 \pi} |b'(k_F R)|^2
= \left( \frac{1}{2 Z^4}-\frac{1}{2 Z^6}+\frac{5}{4 Z^8}
+ ...\right) \exp{(-2R/\xi)}
+ {\cal O}\left(\exp{(-4R/\xi)}\right)
\end{eqnarray}
at $\omega=0$.
We deduce the first non vanishing term in the large-$R$,
large-$Z$ expansion of the non local transmission:
\begin{eqnarray}
\label{eq:Tprime}
T'&=&\int_0^{2\pi} \frac{d (k_F R)}{2 \pi} \left(|a'(k_F R)|^2
-|b'(k_F R)|^2\right)\\
&=&- \frac{3}{4 Z^8}\exp{(-2 R/\xi)} + 
{\cal O}\left(\exp{(-4R/\xi)}\right)
,
\end{eqnarray}
having a sign dominated by elastic cotunneling,
in agreement with the Green's function approach in which the
first term in expansion of the non local transmission appears
at order $T^4 \exp{(-2 R/\xi)}$, where the large-$Z$ normal
transmission coefficient is proportional to $Z^{-2}$ \cite{BTK}.

In the case of highly transparent interfaces corresponding to $Z=0$,
we find no crossed Andreev reflection:
$a'(\omega)=0$, in agreement with the Green's function approach
in Ref. \cite{Melin-Feinberg}. The 
elastic cotunneling transmission coefficient for $Z=0$
is given by
\begin{equation}
|b'(\omega)|^2=\frac{4(\Delta^2-\omega^2)\exp{(-2R/\xi)}}
{3\omega^2 [1-\exp{(-2 R/\xi)}]+
\Delta^2 [1+\exp{(-2 R/\xi)}/2]}
.
\end{equation}

\begin{figure}
\begin{center}
\includegraphics [width=1. \linewidth]{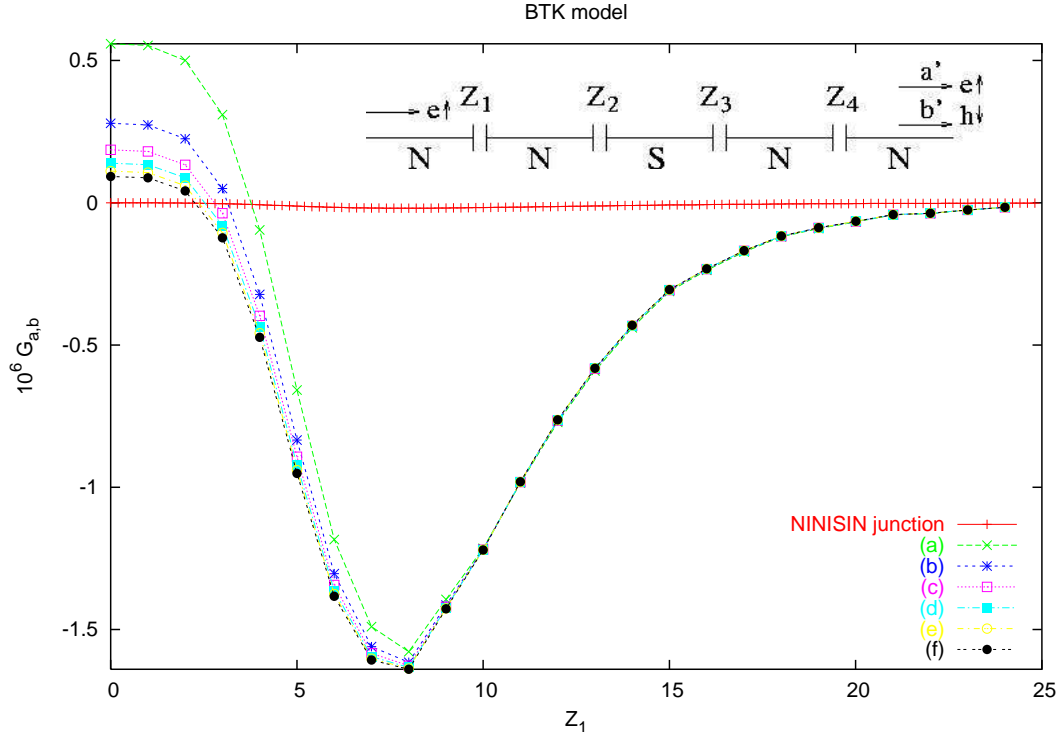}
\end{center}
\caption{(Color online.) Variation of the crossed conductance $G_{a,b}$ (in units of
$e^2/h$) for the junction on Fig.~\ref{fig:schem-btk}c, with
$Z'=Z_1=Z_4$ and $Z=Z_2=Z_3=10$. 
(a) ... (f) correspond to an increasing
values of the precision in the evaluation of the Fermi
phase factors related to the superconductor.  We have also
shown the much smaller
crossed conductance of the NINISIN junction,
as a function of $Z_1$ for the NIN contact, with
the same value of $Z$ for the NIS contacts.
\label{fig:result1}
}
\end{figure}

\section{NINISININ junction}
To describe multiple scattering in the normal electrodes,
we consider now two additional scatterers 
at positions $z_1=-L_1/2$ in the left electrode
and $z_2=L_2/2$ in the right electrode,
described by
the potentials $V'(z)=H' \delta(z-z_1)
+H' \delta(z-z_2)$, and leading to the
barrier
parameter $Z'=2mH'/\hbar^2 k_F$
(see Fig.~\ref{fig:schem-btk}b for the
definitions of $Z$ and $Z'$). We average numerically the non local
transmission coefficient over the Fermi oscillation phases
$\varphi_1=k_F(R-L_1)/2$, $\varphi=k_F R$ and
$\varphi_2=k_F(R-L_2)$. 

The variations of the crossed conductance at zero bias
as a function of $Z'$ for a fixed $Z$
are shown on Fig.~\ref{fig:result1}, as well as the
corresponding crossed conductance for the NINISIN junction.
The integration over the microscopic Fermi oscillation
phases for the latter involves
a double integral so that the accuracy is larger than for the
NINISININ junction involving a triple integral.
As it is visible from the curves (a) - (f) on Fig.~\ref{fig:result1}
corresponding to an increasing precision in the evaluation of the
integrals, the crossed conductance for $Z_1=0$ has
not converged to the limiting value obtained for the NINISIN junction,
meaning that the change of sign in the crossed conductance
at small $Z_1$ 
for the NINISININ junction is an
artifact related to the
lack of precision in the evaluation of the 
triple integral (the crossed conductance at $Z_1=0$ for
the NINISIN junction is indeed negative).
The variation of the crossed conductance
on Fig.~\ref{fig:result1} shows a strong enhancement
by the additional scatterers, suggestive of reflectionless tunneling,
as for a NIS interface \cite{Melsen}.

\section{Green's functions}

\begin{figure}
\begin{center}
\includegraphics [width=.6 \linewidth]{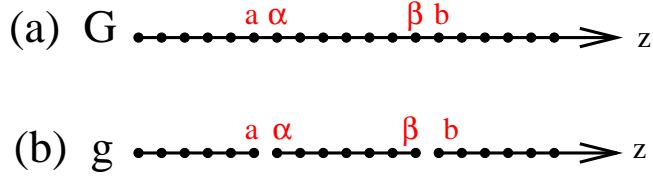}
\end{center}
\caption{(Color online.) The notations used in the evaluation
of the Green's functions of the one dimensional tight-binding model
on the segments $\left[\alpha,\beta\right]$ (b),
from the Green's functions of the full 1D chain (a).
\label{fig:app-1D}
}
\end{figure}
Now we consider the same one dimensional geometry within
Green's functions, and first evaluate
the normal and superconducting
Green's functions with appropriate boundary conditions.
In one dimension,
the Nambu Green's function of a superconductor at distance $R$
and energy $\omega$ is given
by
\begin{equation}
\label{eq:g1D}
\hat{g}(R,\omega)=
\left(\begin{array}{cc} 
g_{1,1}(R,\omega) & g_{1,2}(R,\omega)\\
g_{2,1}(R,\omega) & g_{2,2}(R,\omega)
 \end{array} \right)
,
\end{equation}
with
\begin{eqnarray}
g_{1,1}(R,\omega)&=& \frac{1}{2T}
\left[-\frac{\omega}{s} \cos{(k_F R)}+\sin{(k_F
    R)}\right]e^{-R/\xi(\omega)}\\
g_{2,2}(R,\omega)&=& \frac{1}{2T} \left[
-\frac{\omega}{s} \cos{(k_F R)}-\sin{(k_F R)}\right]e^{-R/\xi(\omega)}\\
g_{1,2}(R,\omega)&=& g_{2,1}(R,\omega)=
\frac{1}{2T} \frac{\Delta}{s} \cos{(k_F R)}e^{-R/\xi(\omega)}
,
\end{eqnarray}
with $s=\sqrt{\Delta^2-\omega^2}$ and $\xi(\omega)=\hbar v_F/s$,
where $T$ is the
bulk hopping amplitude of the one dimensional tight-binding model,
and $v_F$ the Fermi velocity.

The Green's functions on the finite segment $\left[\alpha,\beta
\right]$ can be deduced from Eq.~(\ref{eq:g1D}) by introducing
a self-energy that disconnects the chain \cite{Vecino}.
With the notations on Fig~\ref{fig:app-1D}, we find
\begin{eqnarray}
g_{\alpha,\beta}^{1,1} &=&
\frac{1}{T} \left[1+\frac{1}{4{\cal D}}
\left(1-e^{-4R/\xi(\omega)}\right)\right]\times
\left[-\frac{\omega}{s}\cos{(k_F R)}
+\sin{(k_F R)} \right] e^{-R/\xi(\omega)}\\\nonumber
&&+\frac{1}{2{\cal D}T} \sin{(2k_F R)}
\left[\cos{(k_F R)}+\frac{\omega}{s}\sin{(k_F R)}
\right]e^{-3R/\xi(\omega)}\\
g_{\alpha,\beta}^{1,2} &=&
\frac{1}{T} \left[1+\frac{1}{4{\cal D}}
\left(1-e^{-4R/\xi}\right)\right]\times
\frac{\Delta}{s}\cos{(k_F R)} e^{-R/\xi(\omega)}\\\nonumber
&&-\frac{1}{2{\cal D}T} \sin{(2k_F R)}
\frac{\Delta}{s} \sin{(2 k_F R)} \sin{(k_F R)}
e^{-3R/\xi(\omega)}
,
\end{eqnarray}
with
\begin{equation}
{\cal D} = \frac{1}{4}
\left[1+e^{-4 R/\xi(\omega)}-2 \cos{(2k_F R)} e^{-2R/\xi(\omega)}
\right]
.
\end{equation}
Similar expressions are obtained for $g_{\alpha,\beta}^{2,2}$ and
$\hat{g}_{\alpha,\alpha}$.

\begin{figure}
\begin{center}
\includegraphics [width=1. \linewidth]{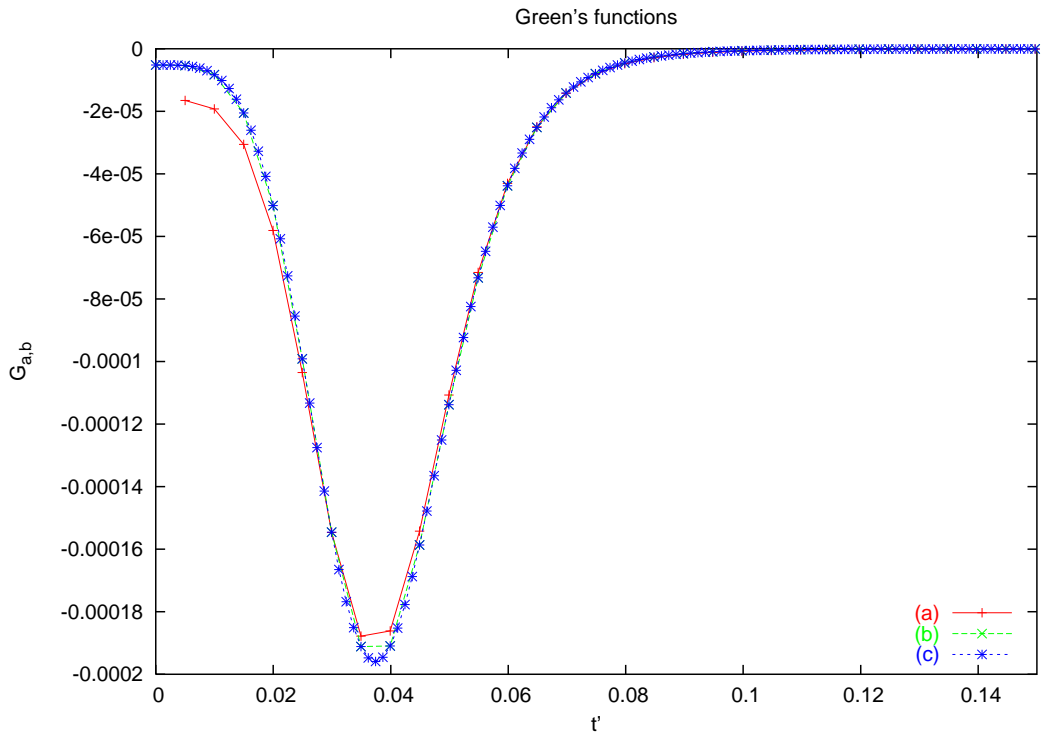}
\end{center}
\caption{(Color online.) Variation of the crossed conductance $G_{a,b}$ (in
units of $e^2/h$) within the
tight-binding model for the junction on Fig.~\ref{fig:schem-btk}d,
as a function of $t'$ for $t/T=0.0499$ (corresponding to $Z=10$
in the BTK model).
The curves (a), (b) and (c) 
correspond to an increasing precision in the
evaluation of the integral.
\label{fig:result2}
}
\end{figure}

Fig.~\ref{fig:result2} shows the Green's functions result
for the variation of the crossed
conductance of the NINISININ junction as a function of $t'$ for a fixed $t$
(see Fig.~\ref{fig:schem-btk}d). The numerical convergence is much faster
than for the corresponding BTK model
calculation because of the reduced dimension
of the matrix to be inverted. We find the same feature as for the BTK
model: the crossed conductance is enhanced by additional scatterers,
as in reflectionless tunneling. Imposing the same normal conductance
in the BTK and in the tight-binding models leads to the relation
\begin{equation}
Z=\frac{1-(t/T)^2}{2 t/T}
,
\end{equation}
leading to a good (but not perfect)
agreement for the crossed conductance
when the tight-binding and BTK results are rescaled on
each other.

\section{Conclusions}

To conclude, we have investigated simple one-dimensional
models consisting of NISIN double interfaces, with additional
scatterers away from the two interfaces, in the spirit of
Ref. \cite{Melsen}. 
We find a strong enhancement of the crossed conductance by the
additional scatterers, suggesting that non local transport at
localized double NIS interfaces
is enhanced by orders of magnitude, like in
reflectionless tunneling at a single NIS interface. The geometry
studied here is such that the Thouless energy associated to the
dimension of the structure parallel to the interfaces is larger
than the bias voltage. This reflectionless tunneling regime 
is not expected to correspond to the experiment in Ref. \cite{Russo}
because of the extended interfaces in this experiment, but
may be probed in future experiments with disordered normal metals
and interfaces of reduced extension.
Finally, we also evaluated the crossed conductance as a function of
energy, and found no sign change when the energy is increased
for the BTK model: the crossed conductance is
dominated by elastic cotunneling at all energies.

\stars

The authors thank D. Feinberg and M. Houzet for helpful discussions
and in particular D. Feinberg 
for participating in the early BTK model calculations for
the NISIN structure.


\begin{thebibliography}{99}
\bibitem[*]{crtbt} U.P.R. 5001 du CNRS, Laboratoire conventionn\'e avec
  l'Universit\'e Joseph Fourier

\bibitem{Andreev} A.F. Andreev, Sov. Phys. JETP {\bf 19}, 1228 (1964).

\bibitem{BTK} G.E. Blonder, M. Tinkham, and T.M. Klapwijk,
Phys. Rev. B 25, 4515 (1982).

\bibitem{Hekking-Nazarov} F.W.J. Hekking and Yu.V. Nazarov
Phys. Rev. Lett. {\bf 71}, 1625 (1993);
Phys. Rev. B {\bf 49}, 6847 (1994).

\bibitem{Beenakker-revue} C.W.J. Beenakker
Rev. Mod. Phys. {\bf 69}, 731 (1997).

\bibitem{Melsen} J.A. Melsen and C.W.J. Beenakker,
Physica (Amsterdam) {\bf 203B}, 219 (1994).

\bibitem{Lambert} C.J. Lambert and R. Raimondi,
J. Phys.: Condens. Matter {\bf 10}, 901 (1998).
  
\bibitem{Jedema} F.J. Jedema, B. J. van Wees, B. H. Hoving,
A. T. Filip and T. M. Klapwijk,
Phys. Rev. B {\bf 60}, 16549 (1999).

\bibitem{Byers} J. M. Byers and M. E. Flatt\'e,  Phys. Rev. Lett.
{\bf 74}, 306 (1995).
  
\bibitem{Deutscher} G. Deutscher and D. Feinberg,
App. Phys. Lett. {\bf 76}, 487 (2000).
  
\bibitem{Samuelsson} P. Samuelsson, E. V. Sukhorukov, and M.
B\"uttiker, Phys. Rev. Lett. {\bf 91}, 157002 (2003);
D. Sanchez, R. Lopez, P. Samuelsson
and M. Buttiker, Phys. Rev. B {\bf 68}, 214501 (2003).

\bibitem{Prada} E. Prada and F. Sols, Eur. Phys. J. B
\textbf{40}, 379 (2004).

\bibitem{Koltai} C. J. Lambert, J. Koltai, and J. Cserti,
{\it in Towards the Controllable Quantum States, Mesoscopic Superconductivity
and Spintronics}, Eds H. Takayanagi and J. Nitta, World Scientific
(2003).

\bibitem{Falci} G. Falci, D. Feinberg, and F.W.J. Hekking,
  Europhys. Lett. {\bf 54}, 255 (2001).

\bibitem{Melin-Peysson} R. M\'elin and S. Peysson,
Phys. Rev. B \textbf{68}, 174515 (2003); R. M\'elin,
Phys. Rev. B {\bf 72}, 134508 (2005).

\bibitem{Chte} N.M. Chtchelkatchev, I.S. Burmistrov, Phys. Rev. B
{\bf 68}, 140501 (2003).

\bibitem{japs} T. Yamashita, S. Takahashi and
S. Maekawa, Phys. Rev. B {\bf 68}, 174504 (2003);
Phys. Rev. B {\bf 67}, 094515 (2003).

\bibitem{Melin-Feinberg} R. M\'elin and D. Feinberg,
Phys. Rev. B {\bf 70}, 174509 (2004).

\bibitem{Melin} R. M\'elin, cond-mat/0510837 (2005).

\bibitem{Tadei} F. Taddei and R. Fazio,
Phys. Rev. B {\bf 65}, 134522 (2002).

\bibitem{Pistol} G. Bignon, M. Houzet, F. Pistolesi, and
F. W. J. Hekking, Europhys. Lett. {\bf 67}, 110 (2004).

\bibitem{Beckmann} D. Beckmann, H. B. Weber, and H. v. L\"ohneysen
Phys. Rev. Lett. {\bf 93}, 197003 (2004);
D. Beckmann and H. v. L\"ohneysen, LT 24 conference proceedings,
cond-mat/0512445 (2005).

\bibitem{Russo} S. Russo, M. Kroug, T. M. Klapwijk, and A. F. Morpurgo
Phys. Rev. Lett. {\bf 95}, 027002 (2005).


\bibitem{Vecino} E. Vecino, A. Mart\'in-Rodero, and A. Levy Yeyati
Phys. Rev. B {\bf 64}, 184502 (2001).

\end{thebibliography}
\end{document}